# A dynamical population modeling of invasive species with reference to the crayfish Procambarus Clarkii

**Short title: Invasive crayfish model**

Gianluca Martelloni[1], Franco Bagnoli[1,3], Stefano Marsili Libelli[2]

[1] Department of Energy Engineering and CSDC, University of Firenze, Firenze (IT), via S. Marta 3 50139 Firenze (Italy), telephone: +39 055 4796592, email: gianluca.martelloni@unifi.it

[2] Department of Systems and Informatics, University of Firenze, Firenze (IT)

[3] Also INFN, sez. Firenze

## Abstract

In this paper we present a discrete dynamical population modeling of invasive species, with reference to the swamp crayfish *Procambarus clarkii*. Since this species can cause environmental damage of various kinds, it is necessary to evaluate its expected in not yet infested areas. A structured discrete model is built, taking into account all biological information we were able to find, including the environmental variability implemented by means of stochastic parameters (coefficients of fertility, death, etc.). This model is based on a structure with 7 age classes, *i.e.* a Leslie mathematical population modeling type and it is calibrated with laboratory data provided by the Department of Evolutionary Biology (DEB) of Florence (Italy). The model presents many interesting aspects: the population has a high initial growth, then it stabilizes similarly to the logistic growth, but then it exhibits oscillations (a kind of limit-cycle attractor in the phase plane). The sensitivity analysis shows a good resilience of the model and, for low values of reproductive female fraction, the fluctuations may eventually lead to the extinction of the species: this fact might be exploited as a controlling factor. Moreover, the probability of extinction is valuated with an inverse Gaussian that indicates a high resilience of the species, confirmed by experimental data and field observation: this species has diffused in Italy since 1989 and it has shown a natural tendency to grow. Finally, the spatial mobility is introduced in the model, simulating the movement of the crayfishes in a virtual lake of elliptical form by means of simple cinematic rules encouraging the movement towards the banks of the catchment (as it happens in reality) while a random walk is imposed when the banks are reached.

# Introduction

The subject of this work is a dynamical model of population growth for the freshwater crayfish *P. clarkii*. It is important to evaluate the impact of this crayfish in not yet infested sites, since *P. Clarkii* may cause a variety of environmental damage, including the instability of the river and basin banks due to the activity of digging to construct hunts (shelter for the winter, for the period of molt and reproduction) [1, 2, 3, 4] and damage to agriculture (its food consists of small animals, shoots and plant debris). Moreover, the crayfish activity increases the water turbidity, with consequent reduction of sunlight penetration that decreases the plant production [3]. *P. Clarkii* belongs to the Arthropoda phylum, Crustacea class and Decapoda order. It is native to Central and South America and was imported into Europe (Spain the first nation) in 1972 [5]. The species was subsequently introduced in other European countries, including Portugal, England, France, Belgium, Germany, Austria, the Netherlands [6] and Switzerland [7]. In Italy the first breeding population was found in Piedmont region in 1989, after the escape of specimens from an experimental farm. Since 1990, *P. clarkii* has been found in many ponds and streams of several provinces of northern and central Italy [8], where populations seem to increase rapidly, in contrast to the native species, *Austropotamobius pallipes*. The individuals of this species can reach 12.5 cm in total length, have a high adaptability to the environment (for example to the haunt conditions) and resistance to high or low temperature and to diseases; they can also resist several hours (up to 24) out of water, moving on the ground near the banks or commuting to other streams. Moreover, they have a high reproductive capacity, laying about 300-500 eggs and occasionally up to 600 eggs for the females of larger size. *P. clarkii* is a *r*-selected species, it has a reproductive strategy typical of colonizers and of species that live in unstable environments, in other words it has the potential to rapidly multiply, producing a very large offspring at the beginning of its life cycle. This is certainly advantageous in environments subject to a short duration, allowing organisms to quickly colonize new habitats and exploit new resources. The *r*-selection species spend the greatest part of their life in an almost exponential phase, in according to *r* growth ratio:

$$r = \frac{1}{N} \cdot \frac{dN}{dt} \qquad (1)$$

where N is the number of individuals and *t* is the time. The life cycle of this species is highly plastic according to the geographical area, mainly in relation to the hydrological cycle and water temperature [9]. Differences between regions were found in the starting and duration of the storage period spent in the haunt (hibernation/aestivation), and in the number of reproductive events. A study of an Italian population [8], showed that the period of hibernation in the haunts stretches throughout the winter (November to March), after which two reproductive events take place. The presence of breeding populations in parts of central and northern Italy [7] suggests that low temperatures are not, as reported in the literature, a factor that limits their reproductive success and the distribution of the species. Laboratory experiments have also shown the shrimp ability to survive and maintain high growth rates at temperatures higher than those tolerated by the native crayfish (Austropotamobius pallipes italicus). All these features make this animal a winning species.

In our model the crayfish population is divided into seven age classes [10]: this choice allows to fully explain the dynamic population growth considering the fertility, the death rate of each age class and the mortality of eggs that are preyed on females.

The real dynamics of *P. clarkii* is rather articulated for what concerns its reproductive biology and the parameters that influence the individual growth. Moreover, the reproductive dynamics is not a unique event but rather is distributed on a period, and therefore it influences the class transitions of individuals. For this reason, we estimate that a discrete model is more flexible and more controllable. In the following we describe the details of the population model that has been calibrated with the experimental laboratory derived parameters (death rate, fertility rate, etc.) provided by DEB.

Moreover in this manuscript we show a simple mobility model, simulating the crayfish invasion of a virtual lake of elliptical form, based on simple cinematic rules. In this implementation the population and the mobility model are not integrated, but independent one from the other, *i.e*, the individuals, generated by population model, obey the rules of the mobility model.

## The population dynamics modeling

In this section we show the model that we have developed, *i.e.*, a discrete one based on age classes. The population is divided into seven age groups, the first is called Newborn, the next three Yearlings and the last three Adults. This choice is suggested by literature in which *P. Clarkii* is always divided into various classes of size or age. Each age group has an average duration of 80 days, as suggested by Anastacio [10], after which it passes into the next class.

Each class has its own population or mortality coefficient that takes into account any intra and inter-specific predations. These parameters should decrease with class advancements, as shown in [10]. Moreover, as will be seen in more detail subsequently, these indexes depend on the water temperature and water level.

The reproductive maturity is reached in the 3rd class, *i.e.*, in about 3-5 months of age [11]: so the last five classes participate in the breeding. Individuals in these classes has already completed the maturity phase, the length of their cephalothorax (head plus thoracic segments) is 60 mm in average [12]. We introduce a growth function of the cephalothorax (CEF) in which the coefficient of intrinsic growth is temperature dependent. In addition we take into account a correspondence between the length of the cephalothorax and fertility.

The last class of Adults (7$^{th}$ class) has duration of 80 days so that the total life span of a generic individual is about 560 days (18.4 months), which is approximately their life expectancy in nature: in the literature it is reported that while in laboratory animals live more than 4 years, in the wild life expectancy rarely exceeds 12-18 months [13].

The stochastic components of the model concern the height of water, considering the water level of a fictitious basin with a random variation $\Delta$ at each time step, and the length of the cephalothorax in each class since, obviously, not all individuals can be of the same size.

P. Clarkii individuals have two mating periods, one in spring and one in summer [14], not concentrated in a specific time, but each one is distributed over a period of two months with an approximately sinusoidal trend in absolute value (there is a growing phase of the couplings that reach a peak in the middle of the period and then decreases). There is a delay of about 60 days between the coupling and the corresponding birth of Newborns. The fraction of females that hatch at least one egg is taken into account in the calculation of fertility. We also consider the high mortality of eggs, which are often re-eaten by the same females.

For the population dynamics we can refer to the state diagram reported in Figure 1, in which the seven classes representing the population are shown, with indication of the fertility and mortality terms related to each single class, the delays between the mating period and the hatching of the eggs and finally the delays due to the passage, during growth and aging, from a class to the contiguous one.

The time step of the model is 5 days, a compromise between efficiency and accuracy, especially for what concerns the reproductive events. So, for example, the coupling step is distributed over 12 periods. In Figure 2 an example of the intra-dynamics between two contiguous classes is shown. At each time step, the births of Newborn is registered in a shift vector after which the individuals are subjected to a mortality rate. After a delay, Newborns pass to the Yearling class.

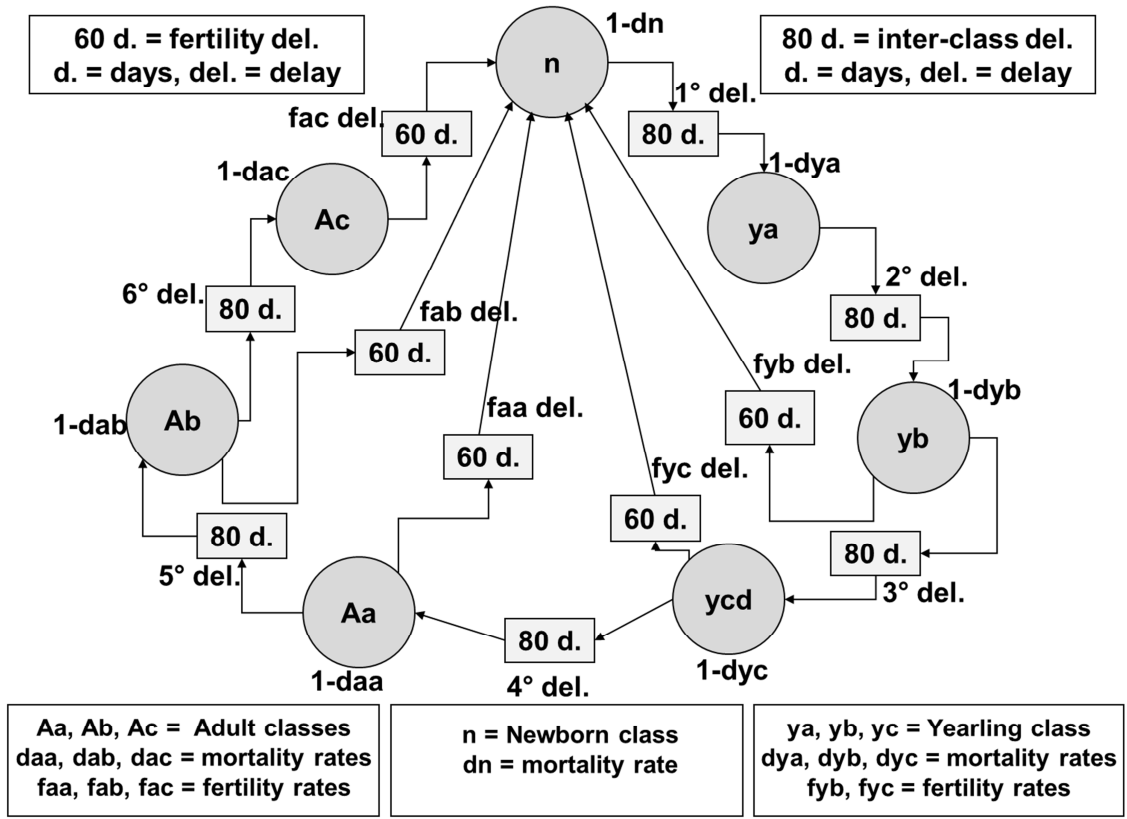

Figure 1: State diagram of the dynamic discrete model based on seven age classes with fertility and inter-class delays.

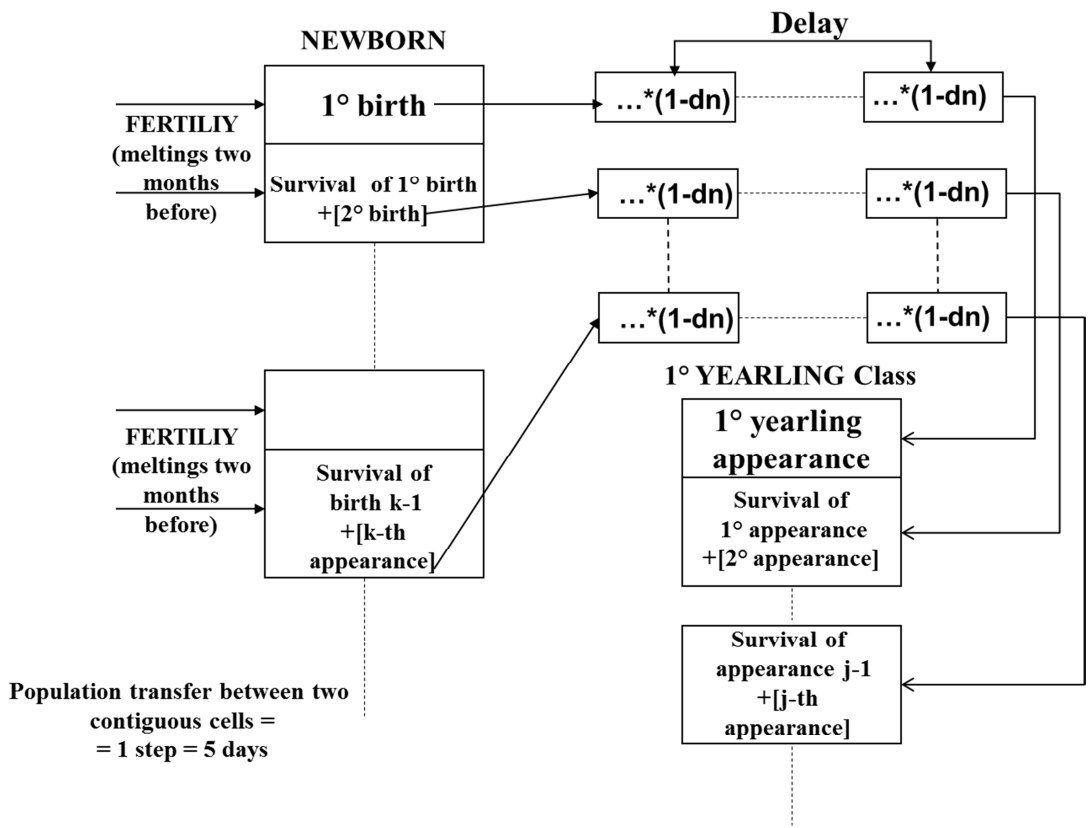

Figure 2: Example of intra-class dynamics between the first and second class.

As mentioned above, the reproductive phase is distributed in two period as shown in Figure 3 [14].

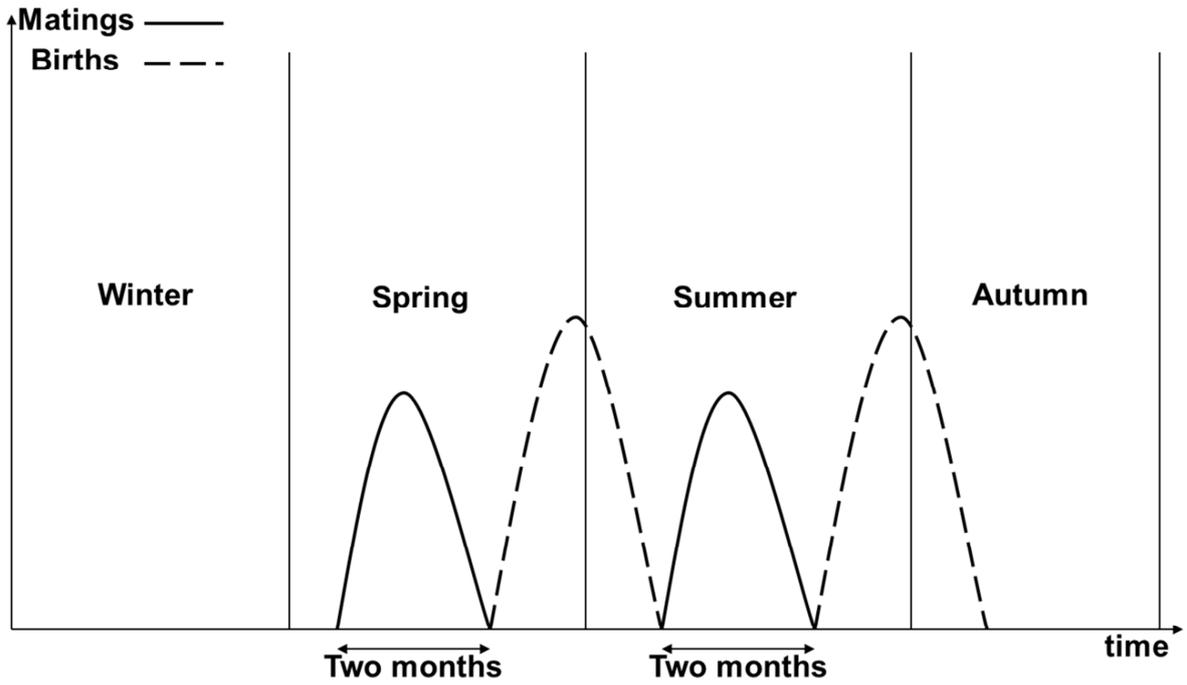

**Figure 3: Mating and birth distribution in time.**

The fraction of female participating to reproduction in each class is approximated by:

$$F_f(t) = (F_\% / F_k) \cdot |\sin(2\pi \cdot (t/24)|  \qquad (2)$$

where $F_\%$ is a laboratory datum of fertility, $F_k$ is the normalization constant ($F_k = \sum_{t=1}^{24} |\sin(\pi \cdot t / 12)|$),

$t$ is the time variable.

In Table 1 the function (2) is reported for each reproductive class.

**Table 1: Female fraction that participates in the reproductive events.**

| | | $1 \leq t_n \leq 12$ |
|---|---|---|
| | $F_f(t_n) = \dfrac{F_\%}{F_k} \cdot \left|\sin(2\pi \cdot (t_n/24))\right|$ <br><br> $F_k = 15.1915$ | *Starting from birth time for each period of reproduction* |
| $F_{yb}$ | $\left(\dfrac{0.47 \cdot 0.55}{15.1915}\right) \cdot \left|\sin(2\pi \cdot (t_n/24))\right|$ | |
| $F_{yc}$ | $\left(\dfrac{0.47 \cdot 0.56}{15.1915}\right) \cdot \left|\sin(2\pi \cdot (t_n/24))\right|$ | |
| $F_{Aa}$ | $\left(\dfrac{0.47 \cdot 0.57}{15.1915}\right) \cdot \left|\sin(2\pi \cdot (t_n/24))\right|$ | |
| $F_{Ab}$ | $\left(\dfrac{0.47 \cdot 0.58}{15.1915}\right) \cdot \left|\sin(2\pi \cdot (t_n/24))\right|$ | |
| $F_{Ac}$ | $\left(\dfrac{0.47 \cdot 0.59}{15.1915}\right) \cdot \left|\sin(2\pi \cdot (t_n/24))\right|$ | |

Concerning the fertility, the latter is a function of the crayfish size and in our modeling we choose the Noblitt empiric relation [15] between eggs $E_g$ and cephalothorax length $L_t$,

$$E_g = 20.5 L_t - 573.5 \qquad (3)$$

and

$$L_t = L_{MAX} \cdot \left(1 - \exp\left(-\left(d \cdot K \cdot t + (d \cdot C \cdot K / 2\pi) \cdot \sin(2\pi \cdot t / 73)\right)\right)\right)^{1/d} \qquad (4)$$

that represents a saturation curve (Figure 4), where $L_{max}$ is the maximum CEF length (60 mm), $C$ is a parameter between 0 and 1, $d$ represents the deviation from Von Bertalanffy law of which the latter equation represents a modification [16], $K$ is the intrinsic growth rate, which generally is considered constant [10], but in our model it is expressed as a function of the daily temperature,

$$K(T(t)) = \dfrac{K_1}{K_0 - \left(\dfrac{T(t)}{T_{MAX}}\right)^2} \qquad (5)$$

where $K_0$ and $K_1$ are constants, $T_{max}$ is the maximm water temperature and $T(t)$ is the water temperature as function of time $t$.

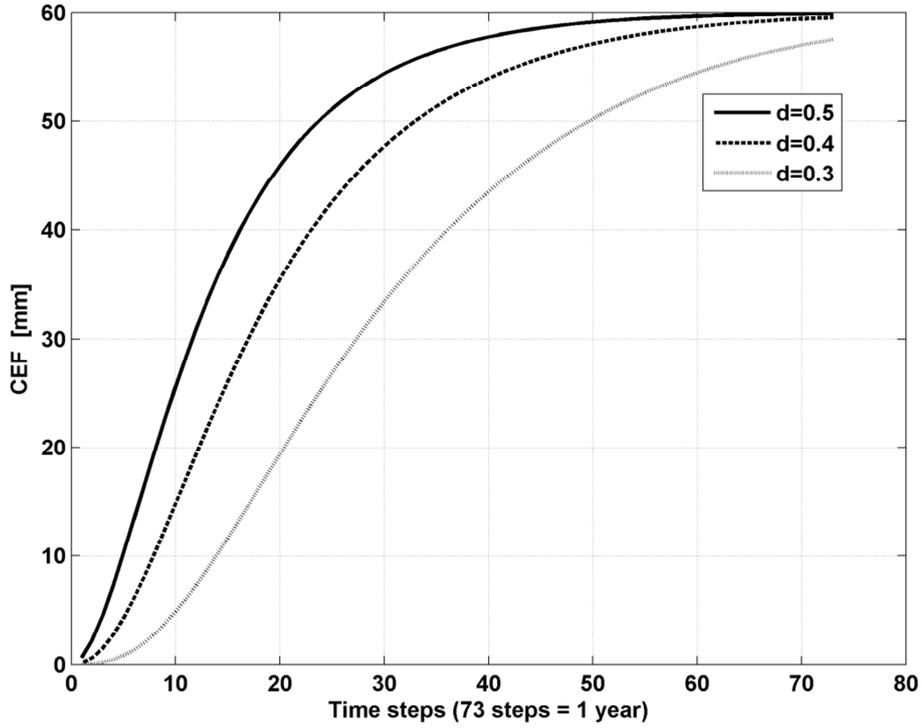

**Figure 4: Cephalothorax (CEF) length growth [mm] in the time.**

In Table 2 the CEF growth law for each class are reported.

**Table 2: Function of CEF growth for each class, from 4° class the maturity is reached.**

| State Variables | Function of CEF growth | Parameters |
|---|---|---|
| $n$ | $L_t = L_{MAX}(1 - \exp(-(dKt + (dCK/2\pi)\sin(2\pi t/73))))^{1/d}$ $K = K(T(t))$ $K(T(t)) = \dfrac{K_1}{K_0 - \left(\dfrac{T(t)}{T_{MAX}}\right)^2}$ | $L_{MAX}=60$ |
| $y_a$ | | $C=0.5$ |
| $y_b$ | | $K_1=0.3$ |
| | | $d=0.25$ |
| $y_c$ | $L_t = L_{MAX} + \sigma \cdot randn(1)$ | $\sigma=1.5$, $randn(1) =$ normal pseudo-number generator |
| $A_a$ | $L_t = L_{MAX} + \sigma \cdot randn(1)$ | |
| $A_b$ | $L_t = L_{MAX} + \sigma \cdot randn(1)$ | |
| $A_c$ | $L_t = L_{MAX} + \sigma \cdot randn(1)$ | |
| $E_g = 20.5 L_t - 573.5$, $L_t$ [m]; Noblitt relation [14] **Valid from the 3° class ($y_b$)** | | |

The mortality rates are function of temperature and water level, according to the following relationship, valid for a generic class,

$$D(T,H) = D_{max} \cdot \exp\left( \log\left(\frac{D_{min}}{D_{max}}\right) \cdot \frac{T}{T_{max}} \cdot \frac{H}{H_{max}} \right) \tag{6}$$

where $D_{max}$ and $D_{min}$ are respectively the maximum and minimum mortality coefficient, $T_{max}$ and $H_{max}$ are respectively the maximum temperature and water level, while $T$ and $H$ are temperature and water level (Figure 5).

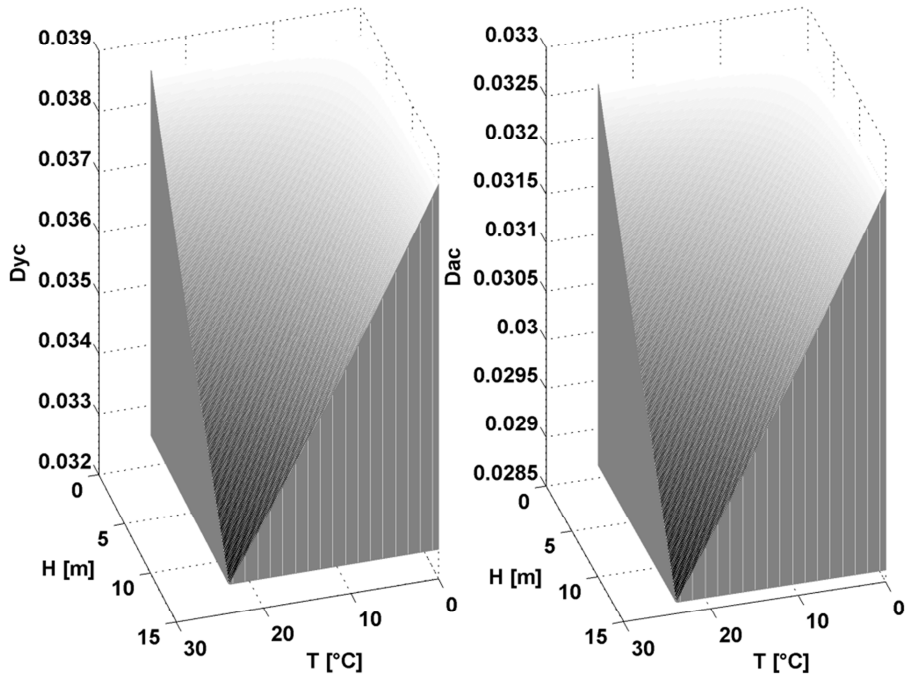

Figure 5: The mortality coefficients for 3° class Yearlings (Dyc) and 3° class Adults (Dac).

Concerning the mortality coefficients $D_{max}$ and $D_{min}$, the latters are deduced by the laboratory experiments where the data are calculated on a monthly base. Since the time step is 5 days, the re-calculated mortality rates are

$$D_5 = 1 - \sqrt[6]{1 - D_{30}} \tag{7}$$

where $D_5$ is a mortality coefficient for 5 days, while $D_{30}$ is the monthly one (6 time steps). The mortality rates for each age classes used in ours model are reported In Table 3. The laboratory data, in terms of the mortality coefficients provided by DEB, are average or maximum values; therefore the minimum coefficients are deduced by inverting the average formula (see Table 3).

Moreover, in Table 4 we show the value of the temperatures and water levels of the fictitious basin adopted in our model, in particular the water temperature is that typical of the Central Italy latitude. The fictitious water levels are summed with small stochastic variations, as indicated in Table 4, to take into account the natural variations of the lake environment (rainfall, drought, etc.).

**Table 3: Mortality functions for each class.**

| $D(T,H) = D_{max} \cdot \exp\left(\log\left(\frac{D_{min}}{D_{max}}\right) \cdot \frac{T}{T_{max}} \cdot \frac{H}{H_{max}}\right)$ | | $T_{max} = 24°C$ | $H_{max} = 21m$ |
|---|---|---|---|
| Classes | $D_{mean}$ | $D_{max}$ | $D_{min}$ |
| $n$ | $1-(1-0.217)^{1/6}$ | $1-(1-0.26)^{1/6}$ | |
| $y_a$ | $1-(1-0.2111)^{1/6}$ | $1-(1-0.2525)^{1/6}$ | |
| $y_b$ | $1-(1-0.2052)^{1/6}$ | $1-(1-0.245)^{1/6}$ | $D_{mean} = \frac{D_{max}+D_{min}}{2}$ |
| $y_c$ | $1-(1-0.1935)^{1/6}$ | $1-(1-0.23)^{1/6}$ | $\Downarrow$ |
| $A_a$ | $1-(1-0.1818)^{1/6}$ | $1-(1-0.215)^{1/6}$ | $D_{min} = 2 \cdot D_{mean} - D_{max}$ |
| $A_b$ | $1-(1-0.1759)^{1/6}$ | $1-(1-0.2075)^{1/6}$ | |
| $A_c$ | $1-(1-0.17)^{1/6}$ | $1-(1-0.20)^{1/6}$ | |

**Table 4: Average temperature and water level of the fictitious basin.**

| Months | Average temperature (°C) | Water Levels (m) | |
|---|---|---|---|
| January | 2 | $16 + \sigma$ | |
| February | 6 | $18 + \sigma$ | |
| March | 10 | $19 + \sigma$ | |
| April | 14 | $21 + \sigma$ | $\sigma = (H_{mean}/H_{max}) \cdot randn(1)$ |
| May | 18 | $16 + \sigma$ | |
| June | 22 | $14 + \sigma$ | |
| July | 24 | $10 + \sigma$ | |
| August | 23.5 | $7 + \sigma$ | |
| September | 21 | $9 + \sigma$ | |
| October | 16 | $10 + \sigma$ | |
| November | 9 | $13 + \sigma$ | |
| December | 4 | $15 + \sigma$ | |

Let

$$[C] = [C_1,\ldots,C_i,\ldots,C_7] = [n, y_a, y_b, y_c, A_a, A_b, A_c] \text{ and } [t_r, t_b, t_s, t_p, t_d]$$

the 7 age classes of our model and, respectively, the reproduction time, birth time, development of first class time, passing to the next class time and the delay passing to the next class (80 days) time. The equations relative to Newborn class are:

$$\begin{cases} t_r = 1 \ldots t_r \ldots t_{r\max} \\ t_b = 1 \ldots t_b \ldots t_{b\max} \\ t_s = 1 \ldots t_s \ldots t_{s\max} \\ t_p = 1 \ldots t_p \ldots t_{p\max} \\ t_d = 1 \ldots t_d \ldots t_{d\max} \end{cases} \qquad (8)$$

$$nn_{t_b} = \left[ \sum_{Ci=3}^{Ci=7} b_{Ci} \cdot F_{Ci} \cdot C_i(t_r) \cdot \exp(-c_p \cdot F_{Ci} \cdot C_i(t_r)) \cdot (1 - d_{eggs}) \right] \quad (9)$$

$$[SMR_n] = \begin{bmatrix} nn_1 \cdot (1-d_n) & \cdots & nn_1 \cdot (1-d_n)^{t_{d\max}-1} \\ \cdots & \cdots & \cdots \\ nn_{t_{b\max}} \cdot (1-d_n) & \cdots & nn_{t_{b\max}} \cdot (1-d_n)^{t_{d\max}-1} \end{bmatrix} \quad (10)$$

$$n(t+1) = \begin{cases} nn_{t_b} + n(t)(1-d_n) & \text{if } t = t_b \\ n(t)(1-d_n) & \text{if } t = t_s \\ n(t)(1-d_n) - SMR_n(t_b, t_{d\max}-1)(1-d_n) & \text{if } t = t_p \\ 0 & \text{if } t \neq t_{b,s,p} \end{cases} \quad (11)$$

where $b_{Ci}$ are the fertility coefficients of mature classes, $F_{Ci}$ are the female fractions that hatch at least one egg, $d_{eggs}$ (equal to 0.5) and $d_n$ are respectively the intrinsic mortality coefficients of eggs and Newborns, while $c_p$ is the velocity constant of the exponential term of egg predation by females (cannibalism of eggs). The equation (10) represents the shift-matrix-register of the Newborn fractions that pass to contiguous class in the unit of time.

Similar to Newborns, the second class of Yearlings exhibits the following dynamics:

$$\begin{cases} t_c = 1.....t_c.....t_{c\max} \\ t_s = 1.....t_s.....t_{s\max} \\ t_p = 1.....t_p.....t_{p\max} \\ t_d = 1.....t_d.....t_{d\max} \end{cases} \quad (12)$$

$$yya_{t_c} = SMR_n(t_c, t_{d\max}-1) \cdot (1-d_n) \quad (13)$$

$$[SMR_{ya}] = \begin{bmatrix} yya_1 \cdot (1-d_{ya}) & \cdots & yya_1 \cdot (1-d_{ya})^{t_{d\max}-1} \\ \cdots & \cdots & \cdots \\ yya_{t_{c\max}} \cdot (1-d_{ya}) & \cdots & nn_{t_{c\max}} \cdot (1-d_{ya})^{t_{d\max}-1} \end{bmatrix} \quad (14)$$

$$y_a(t+1) = \begin{cases} yya_{t_c} + y_a(t)(1-d_{ya}) & \text{if } t = t_c \\ y_a(t)(1-d_{ya}) & \text{if } t = t_s \\ y_a(t)(1-d_{ya}) - SMR_{ya}(t_c, t_{d\max}-1)(1-d_{ya}) & \text{if } t = t_p \\ 0 & \text{if } t \neq t_{c,s,p} \end{cases} \quad (15)$$

where $t_c$, $t_s$, $t_p$, $t_d$, are respectively the time of passing in this class, the time of the development, the time of passing to the next class and the delay time of passing to the next class. Note that these different temporal steps are shifted respect to the Newborn ones. Moreover $d_{ya}$ is the mortality coefficient of second class and equation (14) represents the analogous shift-matrix-register.

Concerning the dynamics of the classes from 3$^{rd}$ to 6$^{th}$, it results perfectly equal to second one. Finally the 7$^{th}$ class exhibits some differences. Let $A_b$ and $A_c$ the number of individuals of 6$^{th}$ and 7$^{th}$ classes, $t_c$, $t_{ei}$, $t_{ef}$ the time of passing in the last class, the time of evolution at mortality rate $d_{ac}$ and the time of final evolution at exponential mortality rate (acceleration of mortality process due to the old age). We have:

$$\begin{cases} t_c = 1.....t_c.....t_{c\max} \\ t_{ei} = 1.....t_{ei}.....t_{ei\max} \\ t_{ef} = 1.....t_{ef}.....t_{ef\max} \end{cases} \tag{16}$$

$$A_c(t+1) = \begin{cases} SMR_{Ab}(t_c, t_{d\max} - 1)(1 - d_{ab}) + A_c(t)(1 - d_{ac}) & if \quad t = t_c \\ A_c(t)(1 - d_{ac}) & if \quad t = t_{ei} \\ A_c(t)e^{-cd_{ac}} & if \quad t = t_{ef} \end{cases} \tag{17}$$

where $SMR_{Ab}$ is the shift-matrix-register of 6$^{th}$ class $A_b$, that take into account the individuals that pass into the last class. The velocity exponential coefficient $cd_{ac}$ (equal to 0.6) is calibrated to have the death at 36 months of life.

## Simulations and results of population model

We simulate the model considering the invasion of fictitious lake by crayfish with initial conditions reported in Table 5.

**Table 5: initial conditions as number of individuals.**

| | |
|---|---|
| $n$ | 0 |
| $y_a$ | 0 |
| $y_b$ | 0 |
| $y_c$ | 1200 |
| $A_a$ | 1000 |
| $A_b$ | 0 |
| $A_c$ | 0 |

Nowadays we don't have real data of population trend for each class and the data of egg predation are missing. In our simulation we assign at $c_p$ the value 0.03, the latter being the only one not available, while all other parameters of the model are obtained by laboratory experiments. Considering the evolution of population and merging Yearlings and Adults we obtain three populations whose dynamics (without stochastic variations) exhibit an attractor in the phase plane (Figure 6), while adding the random variables the attractor enlarges (Figure 8). Using empirical relation between CEF and crayfish mass [10], it is possible to plot the total biomass evolution in the time without stochastic variations (Figure 7), and with stochastic ones (Figure 9).

The study of the model is completed with sensitivity analysis of the parameters that are varied between -20% and 20% (Table 6). Such studies are performed on the average of the total Adults population. The model exhibits a good robustness and the highest variations are obtained for the coefficient of the Yearlings mortality: this behavior is obviously due to the fact that the Adults follow directly the Yearlings development.

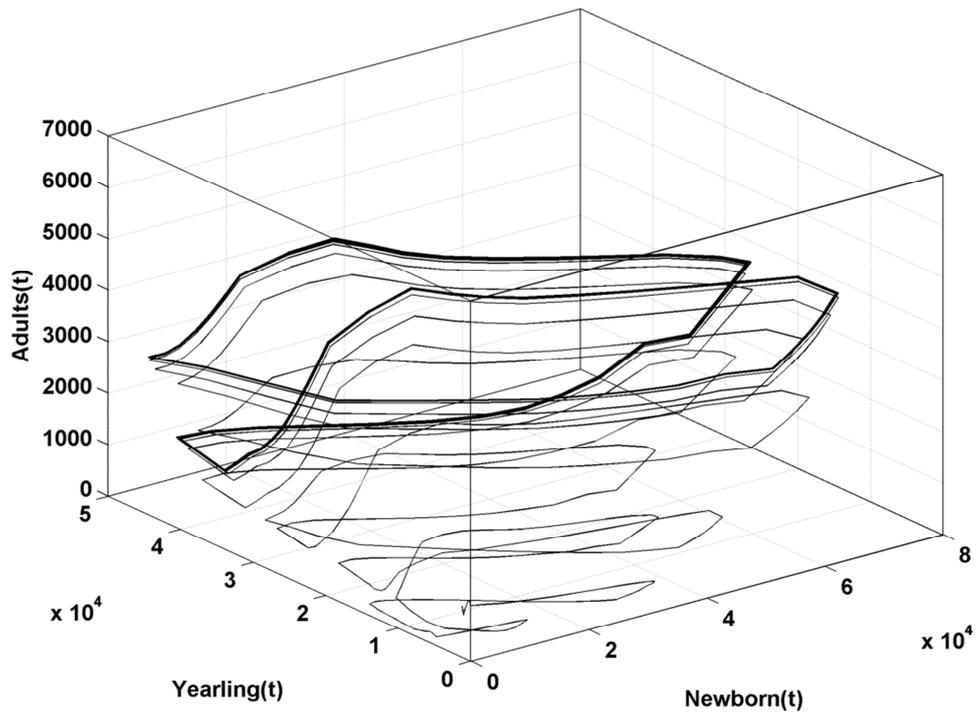

**Figure 6: The three aggregated classes as number of individuals in the phase plane exhibit an attractor (simulation without stochastic variations).**

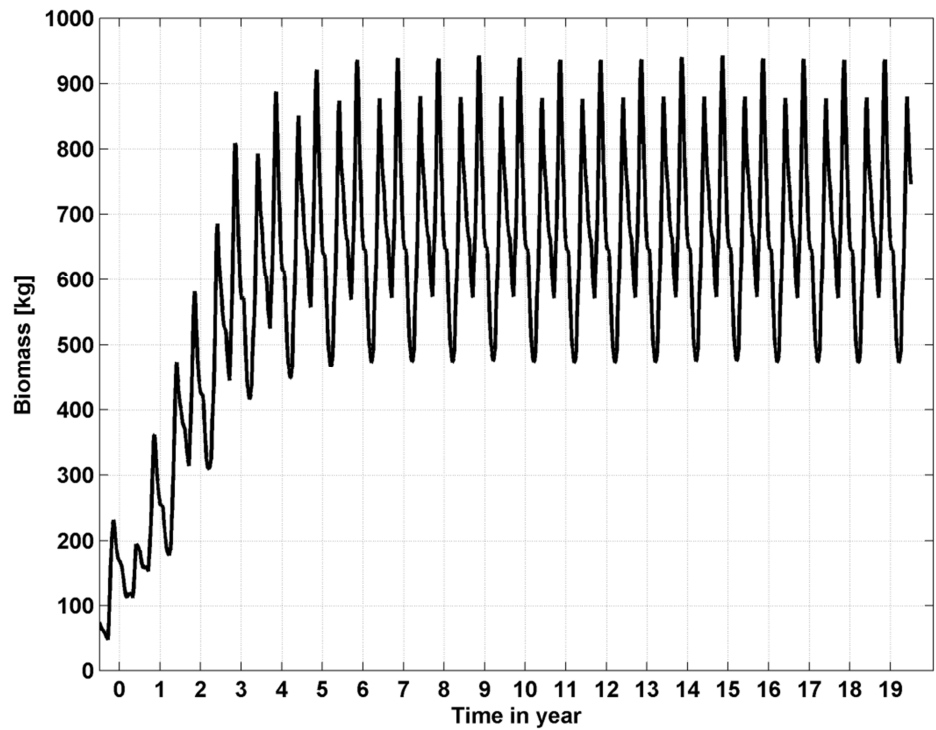

**Figure 7: Total biomass in the time shows a double peak due to the two period of reproduction (simulation without stochastic variations).**

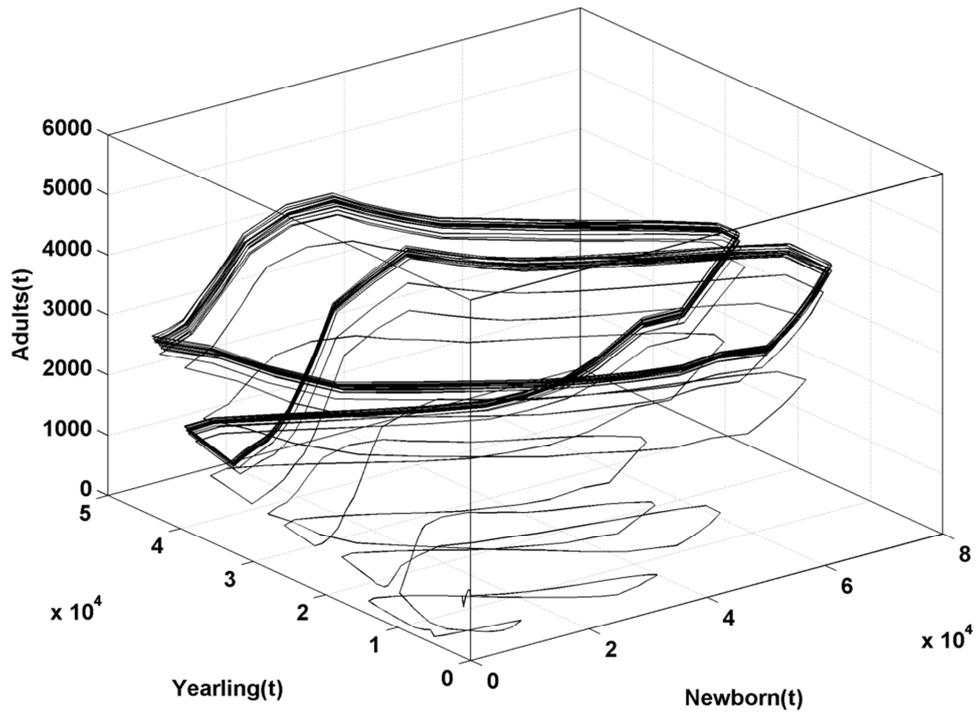

**Figure 8: The three aggregated classes as number of individuals in the phase plane with stochastic variations.**

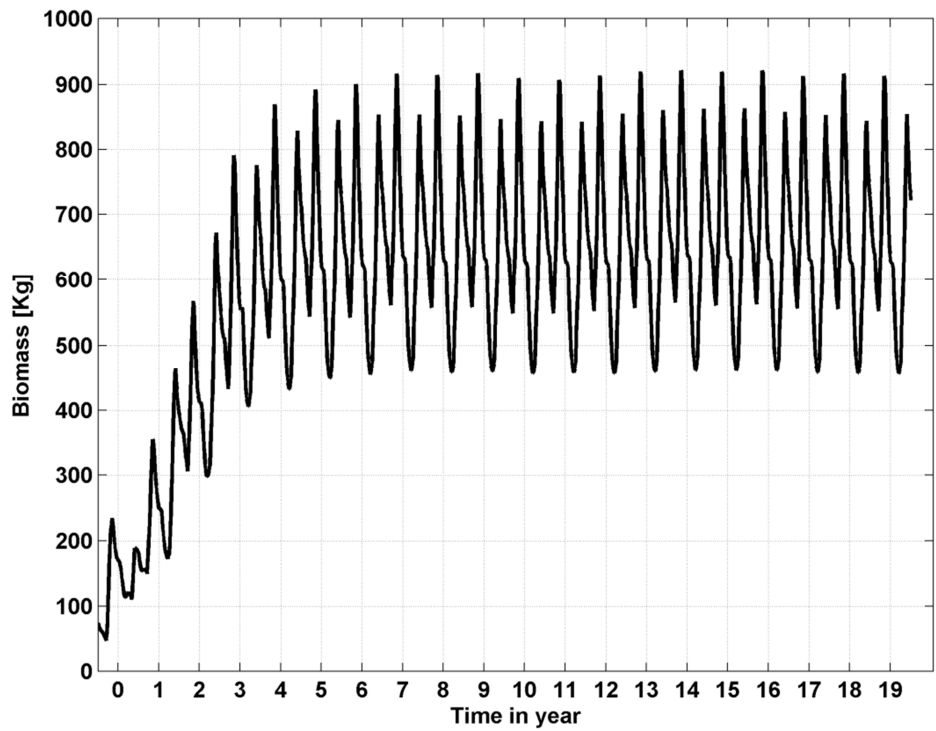

**Figure 9: Total biomass in the time shows a double peak due to the two period of reproduction (simulation with stochastic variations).**

Table 6: Variations in % of the mean of total Adults population to vary the model parameters from -10% to 10% and from -20% to 20%.

| Parameter variations (%) | -10 | 10 | -20 | 20 |
|---|---|---|---|---|
| Model sensitivity (%) | | | | |
| Initial Conditions | -0.45 | 0.98 | -1.11 | 2.02 |
| Fertility | -11.35 | 11.88 | -25.05 | 24.51 |
| Egg Mortality | 11.81 | -12.45 | 24.2 | -24.87 |
| Egg predation | 11.76 | -8.15 | 25.42 | -16.91 |
| Newborns mortality | 9.74 | -8.14 | 18.81 | -16.75 |
| Yearlings mortality | 20.21 | -18.74 | 43.42 | -34.32 |
| Adults mortality | 5.89 | -5.2 | 10.24 | -9.23 |
| Stochastic noise | < 2 | | | |

Moreover we evaluate the extinction probability with the inverse Gaussian distribution method [17, 18]:

$$\begin{cases} G(T_e \mid d, \mu, \sigma^2) = \Phi\left(\frac{-d - \mu T_e}{\sqrt{\sigma^2 T_e}}\right) + \exp\left(\frac{-2\mu d}{\sigma^2}\right) \Phi\left(\frac{-d + \mu T_e}{\sqrt{\sigma^2 T_e}}\right) \\ \Phi(z) = \left(1/\sqrt{2\pi}\right) \int_{-\infty}^{z} \exp\left(-y^2/2\right) dy \end{cases} \quad (18)$$

The latter is applied to the reproductive population (last five classes) where $T_e$ is the estimate time of $G$, $d$ is the logarithm of the initial value of the population at regime, *i.e.*, when transient finished, while the mean $\mu$ and the variance $\sigma^2$ are evaluated by:

$$\begin{cases} \hat{\mu} = \frac{1}{t} \sum_{i=0}^{t-1} \log\left(\frac{Pop_{i+1}}{Pop_i}\right) \\ \hat{\sigma}^2 = \frac{1}{t-1} \sum_{i=0}^{t-1} \left(\log\left(\frac{Pop_{i+1}}{Pop_i}\right) - \hat{\mu}\right) \\ Pop = y_b + y_c + A_a + A_b + A_c \end{cases} \quad (19)$$

The population death evaluation with the latter method shows a low probability of extinction as indicated in Table 7.

The probability density and probability distribution for periods of 200 and 400 years are reported in Figure 10a and 10b, respectively.

**Table 7: Probability of extinction evaluated with inverse Gaussian.**

|          | Probability of extinction |
|----------|---------------------------|
| 5 years  | 0.00002%                  |
| 10 years | 0.00253%                  |
| 15 years | 0.02734%                  |
| 20 years | 0.11468%                  |

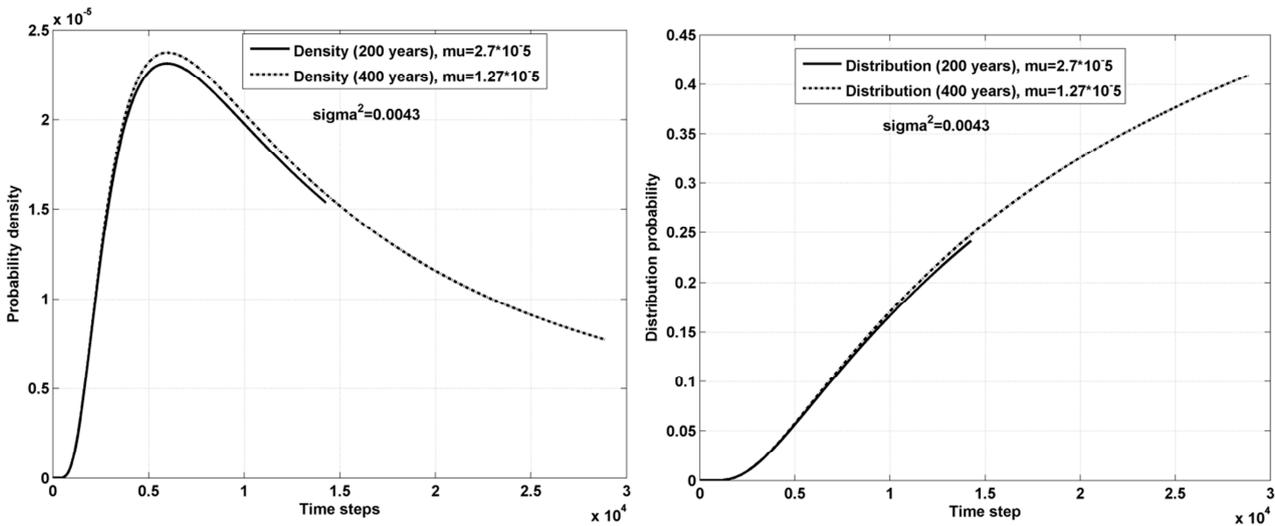

**Figure 10: (a) The probability density of Cox and Miller to estimate the probability extinction, (b) The distribution probability.**

Finally, we simulate a continuous fishing occurring only in the reproductive period with the purpose of controlling the population. In Figure 11 we show a bifurcation diagram, *i.e.*, the asymptotic value of biomass versus the fishing parameters $p_f$ that is simply $(1/p_f) \cdot (population\ value)$, *i.e.*, for example, the value 2 implies to maintain the reproductive classes at the 50% of their value, since is very difficult to fish the young classes (they live principally in the haunt). We have simulated both continuous fishing and a single fishing event. The results are analogous, but the single fishing has to be very intensive to reach the same results of the continuous one. In the diagram of Figure 11 we detect a critical value of the fishing parameter (2.42) above which the population is driven to extinction.

As expected, the system is very sensitive for the critical value of the fishing parameter. After adding stochastic fluctuations, we note that even small variations trigger the survival or extinciton of the species, as shown in Figure 12 where the asymptotic value of the biomass is evaluated for 100 simulations.

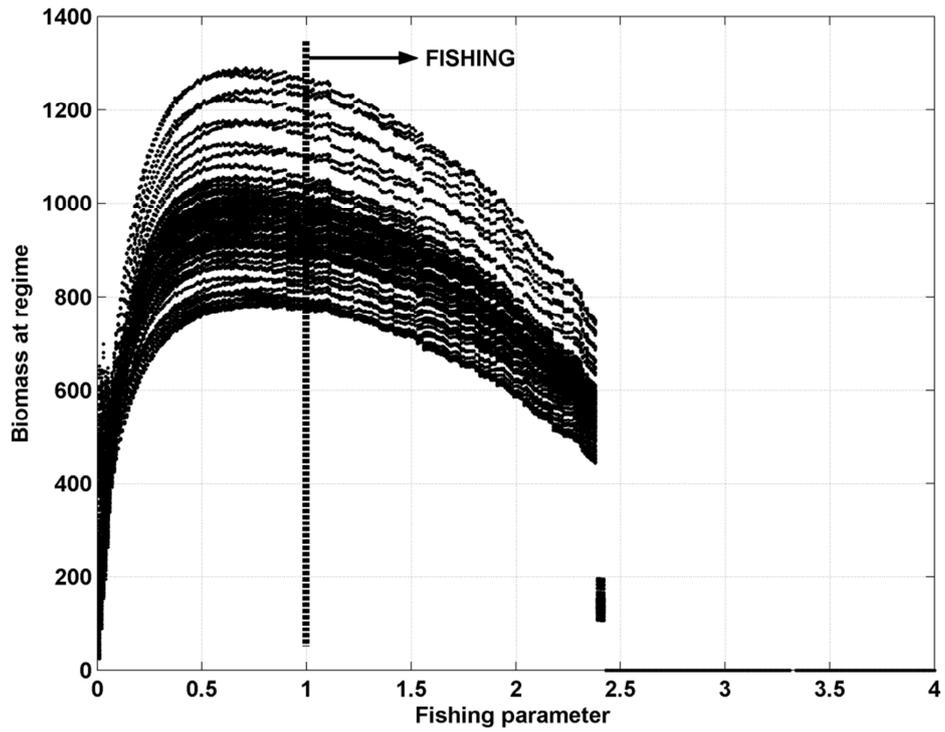

**Figure 11: Bifurcation diagram biomass at regime versus fishing parameter.**

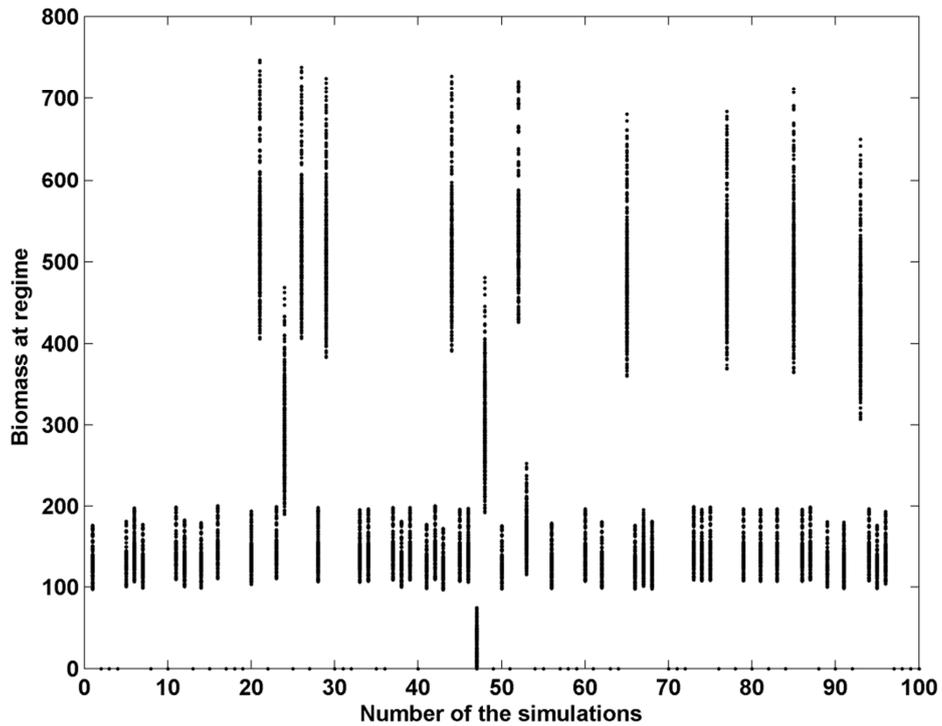

**Figure 12: Asymptotic biomass for the critical value of the fishing parameters for 100 simulations considering stochastic fluctuations.**

# The mobility toy model

Let us introduce the diffusive behavior of the crayfish *P. clarkii* into a single aquatic system, modeled as a virtual lake of elliptical form. This is only a first approximation, additional data are needed to justify a more complex model.

This crustacean exhibit two opposite displacement tendencies: nomadism, where the individuals travel in a large area without well-defined starting or ending points [19], and migration, where movement is directional and the transfer occur in a zone different, from the ecological point of view, of the starting zone [20]. The crayfish excavates hunts, along the banks of the basin, to take refuge in many period of the year (in winter, in dry conditions, during the reproductive period, etc.). The hunts can reach one meter in length, and, in the case of dense populations, the bank is completely full by hunt holes that confer a characteristic hive appearance to it. This crayfish has the capacity of moving out of water with a velocity that varies from 1 to 2 m/d [21]. Then the crayfish density in a small space cab be very high [22].

The mobility and the population model are independent, *i.e.* the crayfish mortality or fertility rate and other parameters of the population evolution are not linked to the crayfish displacements.

The lake is considered a closed system and we start from simple cinematic equations in two dimensions:

$$\frac{d\mathbf{r}}{dt} = \mathbf{v}(\Delta x, \Delta y) \tag{20}$$

where the velocity **v** depends on displacements along *x* and *y* as expressed by equation (20) that is discretized on intervals of 1 m length. Writing explicitly for the two dimensions,

$$\left[\frac{\Delta x}{\Delta t} = v_x(\Delta x, \Delta y), \quad \frac{\Delta y}{\Delta t} = v_y(\Delta x, \Delta y)\right] \tag{21}$$

where $\Delta t = 1\text{d}$ (d = days). The position for each individual at time *t+1* is given by:

$$\left[x_{t+1} = x_t + v_{x,t+1}(\Delta x, \Delta y), \quad y_{t+1} = y_t + v_{y,t+1}(\Delta x, \Delta y)\right] \tag{22}$$

In the equation (22) the velocity term is considered to be a mean value.

We simulate the arriving of crayfish (the initial condition of population modeling) in a lake with a banks of elliptical form. The motion field is shown in Figure (13). The increments on the positions are calculated using a uniform or Gaussian generator of pseudo-random numbers responding to the rules of the scheme of Figure 13, favoring the motion along the banks as it happens in reality.

In Figures 14 and 15 we show the distribution of the occupied positions during one or two simulated years for total Adults population.

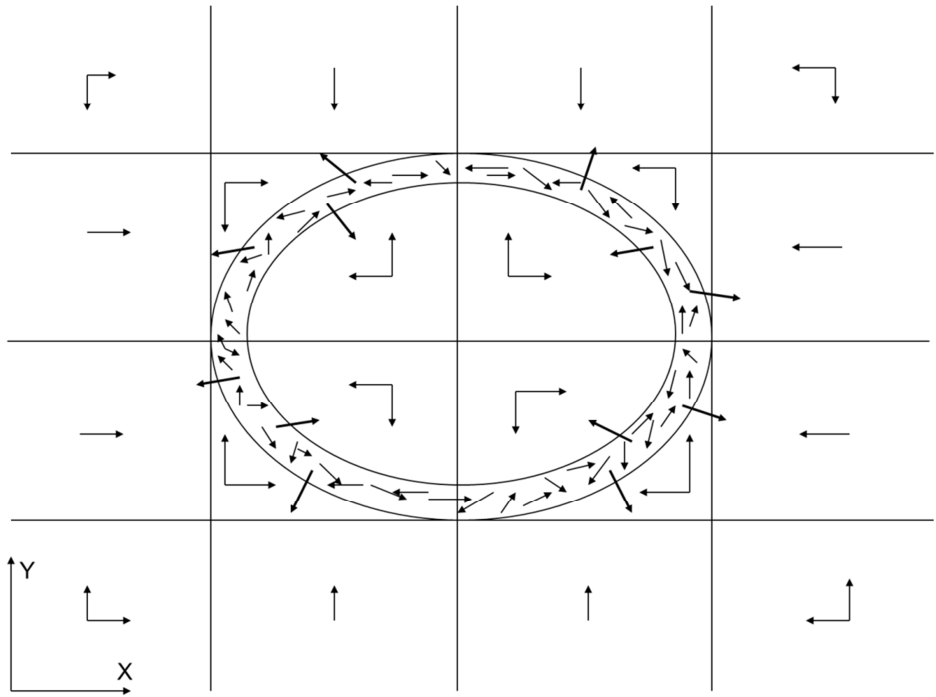

**Figure 13: Virtual lake and the motion field with the assigned movement rules, indicated by the arrows.**

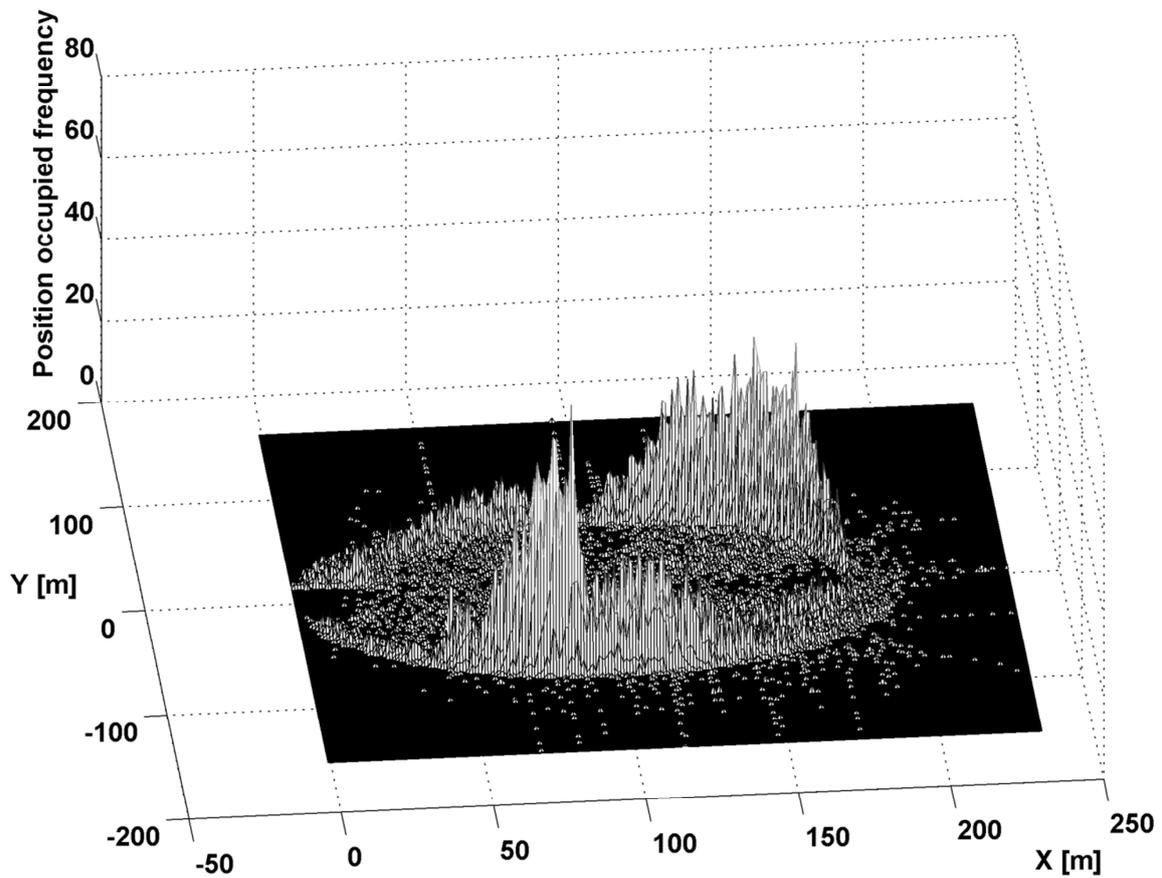

**Figure 14: The frequency of occupied positions simulating one year.**

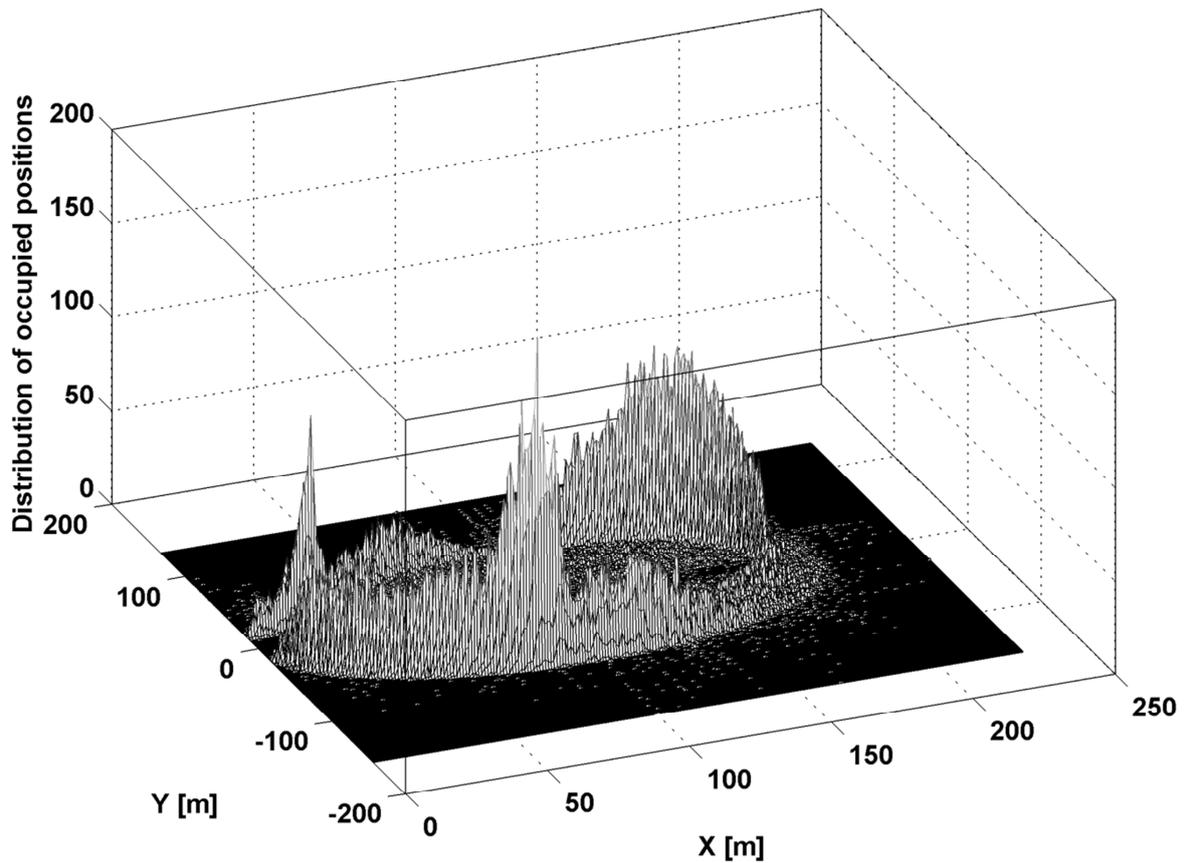

**Figure 15: The frequency of occupied positions simulating two years.**

Then in Figure 16a and 16b we show the occupied position after 5 and 10 days after the arrival of the crayfish in the lake: initially we have a transient represented by a classical diffusion process. In Figure 17 instead the asymptotic spatial pattern is shown. We can observe the colonization of the banks. In this mobility model we consider the displacements of Adults population, and Newborn are generated at random in the lake.

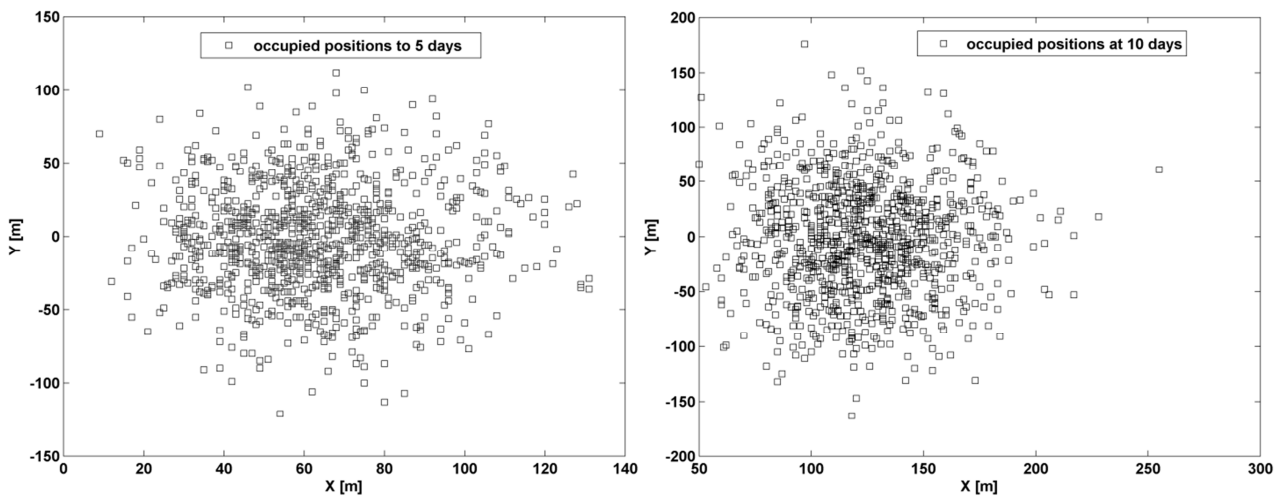

**Figure 16: (a) The occupied positions after 5 days from arriving, (b) After 10 days from arriving.**

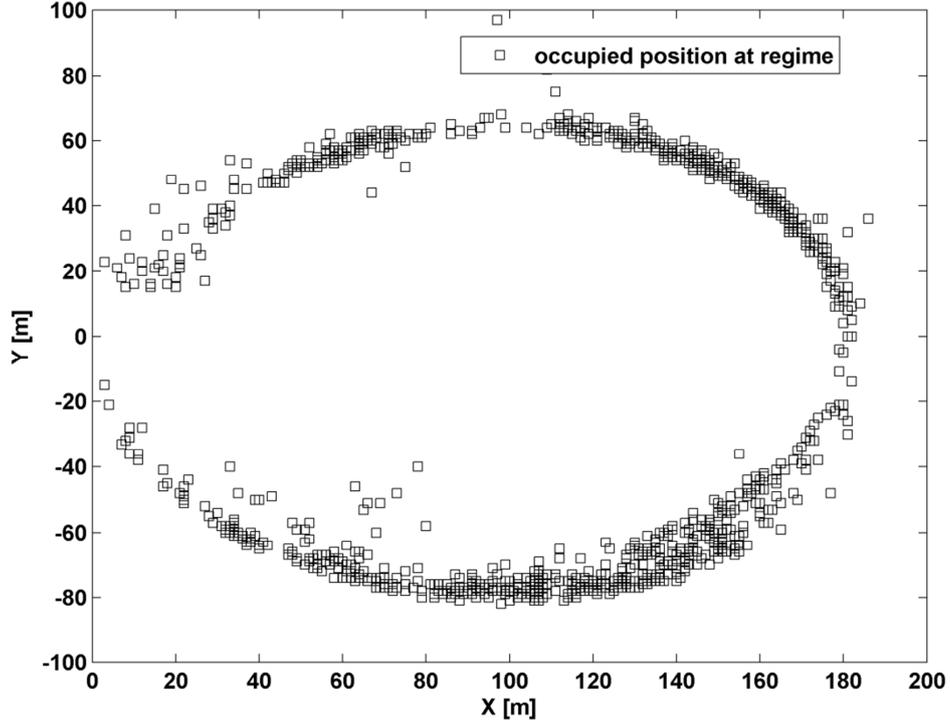

**Figure 17:** Occupied positions at regime in which the banks are colonized by crayfish.

We can analyze the expansion of the invasion, defining the expansion velocities along axis of the lake reference system, as function of the positional field (*i.e.* the positions in the vectors $\mathbf{X}_{t+1}$, $\mathbf{X}_t$, $\mathbf{Y}_{t+1}$, $\mathbf{Y}_t$) as follows:

$$V_x^+ = \frac{\max(\mathbf{X}_{t+1}) - \max(\mathbf{X}_t)}{\Delta t} \quad (23)$$

$$V_x^- = \frac{\min(\mathbf{X}_{t+1}) - \min(\mathbf{X}_t)}{\Delta t} \quad (24)$$

$$V_y^+ = \frac{\max(\mathbf{Y}_{t+1}) - \max(\mathbf{Y}_t)}{\Delta t} \quad (25)$$

$$V_y^- = \frac{\min(\mathbf{Y}_{t+1}) - \min(\mathbf{Y}_t)}{\Delta t} \quad (26)$$

So, for example, Equation (23) indicates the velocity expansion in the positive direction of axis *X*. These velocities exhibits a similar trend, with an initial high velocity for the river flow with subsequent diffusion in the lake and velocity oscillations in the stationary phase due to the displacements along the banks. For instance, the velocity expansion along the positive X direction is reported in Figure (18). The analysis of the position distribution, considering all population, detects a peak density of 15 individuals/m². In literature, in extreme cases, it possible to have a peak density up to 100 individuals/m² [22].

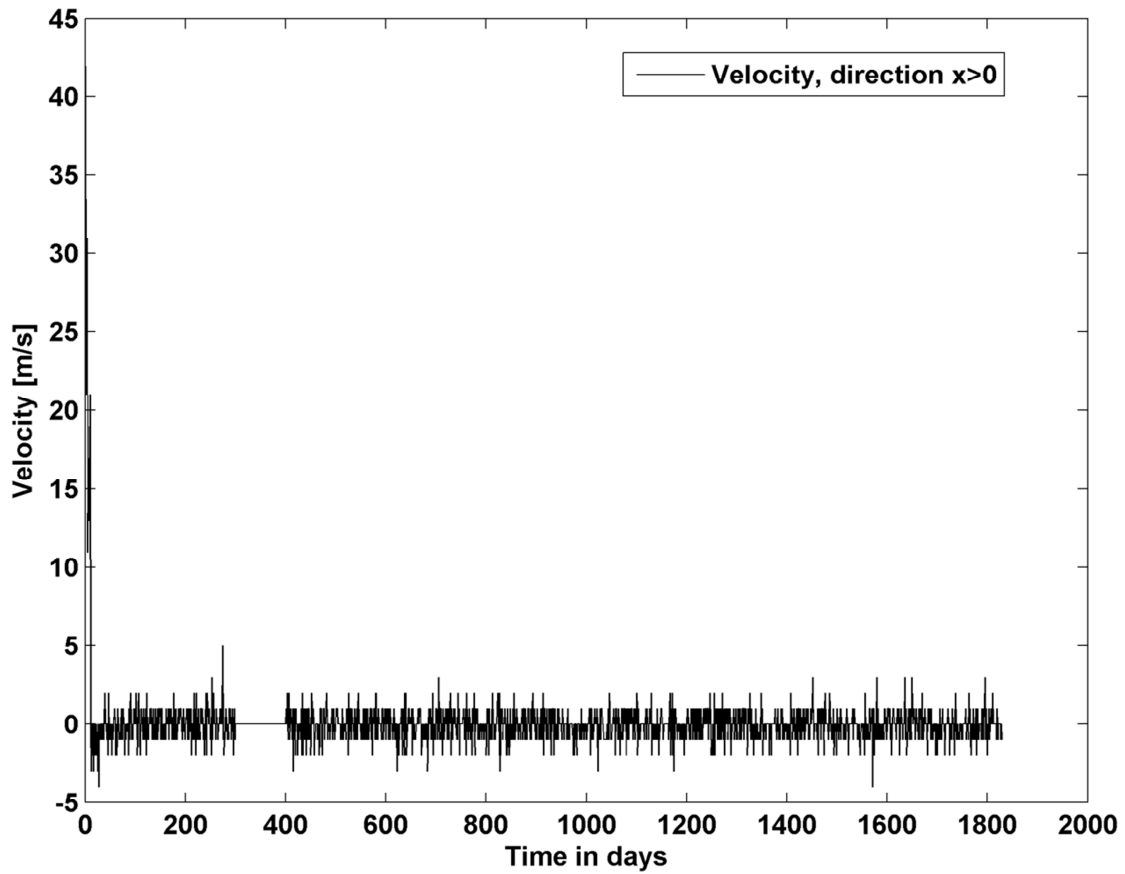

**Figure 18: The velocity expansion of invasion along positive *X* axis direction of the lake reference system.**

## Conclusions

We have modeled the entire life cycle of the shrimp P. Clarkii with a system of time-discrete non-linear equations. The model shows a good robustness in terms of sensitivity to variations at the parameters and exhibits a very low probability of extinction, confirming the experimental observations of the resilience of the species. The values of population density are consistent with data reported in the literature and therefore we estimate that the model is useful for assessing the impact of this invasive species in non-yet-infested environment. We have shown that it is possible to drive the population to extinction after intensive fishing, but clearly more investigations are needed for deciding if this effect is a real one.